\documentstyle[12pt]{article}

\newcommand{\be}{\begin{equation}}
\newcommand{\ee}{\end{equation}}
\newcommand{\al}{\alpha}
\newcommand{\als}{\alpha_s}
\newcommand{\eps}{\varepsilon}
\newcommand{\msbar}{\overline{\rm MS}}
\newcommand{\re}[1]{(\ref{#1})}
\begin{document} 
\thispagestyle{empty}
\thispagestyle{empty}
\begin{flushright}
MZ-TH/99-21\\
May 1999
\end{flushright}
\begin{center}
\vspace*{0.7in}
\hfill 
{\Large \bf
Two-loop corrections to the correlator of tensor 
currents in gluodynamics}
\vspace*{0.5in}

{\large \bf A.A.Pivovarov}\\
\vspace*{0.2in}
{\it Institute for Nuclear Research of the
Russian Academy of Sciences, \\
Moscow 117312}\\
{\rm and}\\
{\it Institut f\"ur Physik, Johannes-Gutenberg-Universit\"at,
  Staudinger Weg 7,\\ 
D-55099 Mainz, Germany}
  \end{center}
\vspace*{0.in}
\centerline{\bf Abstract}
\noindent
Results of evaluating 
the leading order
$\als$ corrections to the correlator of
tensor currents in pure gluodynamics are presented.
These corrections 
to the parton result 
for the correlator are not large numerically that allows one 
to use perturbation theory for the analysis 
of the resonance spectrum within the sum rules
method.

\newpage

Gluonia -- hadronic resonances strongly 
coupled to gauge invariant operators 
built from gluon fields --
are a bright manifestation of the color structure 
of strong interactions. An experimental discovery
of these particles would decisively confirm the validity 
of QCD as a theory of hadrons, (see e.g.~\cite{hagi}).
Theoretically the information on the properties of gluonia can be
obtained (besides numerical 
simulations on the lattice) from the analysis based on 
the sum
rules technique that requires the computation of various correlators of
corresponding gluonic interpolating currents within Operator Product
Expansion (OPE).
Recently an extensive 
review of gluonia properties has been presented in 
ref.~\cite{narison} while 
Borel sum rules and finite energy sum rules 
(FESR) were analyzed earlier in \cite{SVZglu,SVZglu1,FESRZ}
(see also \cite{tenpav,boos0,close,boos,steele}).

One of the most striking features of perturbation theory 
used so far for the analysis of the gluonic current correlators 
within sum rules approach is 
a large magnitude of higher order corrections.
In the case of scalar and pseudoscalar gluonic currents 
the 
next-to-leading order corrections are very large
in the standard $\overline{\rm MS}$-scheme of renormalization
and completely out of control \cite{scaglu}
that makes questionable the applicability of the entire analysis
based on OPE.
Perturbation theory corrections to correlators of gluonic currents
with other quantum numbers have not been available yet.   
In the present letter we fill this gap and report on the results for the
leading order correction to the correlator of 
tensor gluonic currents in pure gluodynamics \cite{diss}.
Tensor mesons in full QCD were considered in \cite{tensh,tennar,tennar1}.

As for the origin of large corrections 
in the scalar and pseudoscalar channels, there are 
strong arguments based on consideration
of instanton contributions possible in both these channels 
that perturbation theory (and OPE) breaks down already 
at very large scale. 
This conclusion was made upon considering the 
magnitude of the power corrections to the gluonic current
correlators 
in the leading order of perturbation theory \cite{SVZglu,SVZglu1}. 
Perturbation theory corrections are also large 
in both channels \cite{scaglu}. On the other hand 
the interaction of instantons 
with tensor gluonic currents is thought to be much weaker (no direct
instanton contribution to the correlator
is possible because of unsuitable quantum numbers 
$J^{PC}=2^{++}$) and one expects that the corrections of perturbation
theory are not extremely large and OPE is still applicable at essentially
smaller momenta than for the (pseudo)scalar case. 
The results of the present note explicitly 
confirm this expectation: 
the computation shows that perturbation theory corrections 
to the correlator of tensor gluonic 
currents are not large and the expansion is valid at smaller scales than in the 
case of the scalar and pseudoscalar gluonic currents.

To analyze the resonance spectrum in the channel of tensor gluonic mesons
one considers a two-point correlator 
of gluonic operators that have a nonvanishing projection 
onto the hadronic state with quantum numbers $J^{PC}=2^{++}$.
The gauge invariant interpolating current for the tensor gluonium 
$j_{\mu\nu}$
is chosen in the following manifest form
\be
j_{\mu\nu}=G^a_{\mu\al}G^a_{\al\nu}+\frac{1}{4}
g_{\mu\nu}G^2 \ , \quad a=1,\ldots,N_c^2-1 
\label{first}
\ee
where 
\[
G_{\mu\nu}=G^a_{\mu\nu}t^a, \quad D_\mu=\partial_\mu-i g_s A_\mu,
\quad A_\mu=A_\mu^a t^a, \quad G_{\mu\nu}=[D_\mu,D_\nu],
\]
$G_{\mu\nu}$ is the gluon field strength,
$A_\mu$ is the gluon field,
$D_\mu$ is the covariant derivative, $t^a$
are 
the generators of the color gauge symmetry group $SU(N_c)$
normalized by the relation
\[ 
{\rm Tr}(t^a t^b)=\frac{1}{2}\delta^{ab}.
\]
Here $G^2$ is a short notation for $G^a_{\mu\nu}G^a_{\mu\nu}\equiv 
\sum_{a,\mu,\nu}G^a_{\mu\nu}G^a_{\mu\nu}$.
The current $j_{\mu\nu}$ coincides with the energy-momentum 
tensor of pure gluodynamics. 
It conserves due to equations of motion for the gluon fields, 
is symmetric and traceless 
at the tree level.
These properties lead to some linear constraints on 
the components of the tensor $j_{\mu\nu}$
that 
effectively kill the superfluous components of the general two-index
tensor in four-dimensional space-time and keep just the necessary
number of components to describe five polarization states of a
massive meson with spin 2 in four-dimensional space-time. 
Radiative corrections is known to destroy this perfect picture 
and lead to the trace anomaly of the form \cite{tra,traa,tranom,tranom1}
\be
j_\mu^\mu= \frac{\beta(\als)}{2\als} G^2
\label{tr1}
\ee
where $\beta(\als)$ is the standard renormalization group
$\beta$-function describing the evolution of the running coupling
constant.
While the rigorous proof of the relation for the trace anomaly
requires rather delicate definitions of the quantities entering 
eq.~\re{tr1},
there is a simple mnemonic rule to recover the proper
normalization of the right hand side of it.
In $D$-dimensional space-time (with $D=4-2\eps$)
one formally finds from eq.~\re{first} with $g_\mu^\mu = D$ 
\be
j_\mu^\mu= \frac{D-4}{4} G^2=-\frac{\eps}{2} G^2 \ .
\label{tr2}
\ee
Recalling that the $D$-dimensional $\beta_\eps$-function is 
given by the expression $\beta_\eps(\als)=-\eps \als + O(\als^2)$
and substituting $\eps=-\beta_\eps(\als)/\als$
into the right hand side of eq.~(\ref{tr2})
one reproduces the correctly normalized 
expression in the right hand
side of eq.~(\ref{tr1}) noticing that 
$\lim_{\eps\rightarrow 0}\beta_\eps(\als)=\beta(\als)$.

The appearance of nonvanishing trace of 
the gluonic operator in eq.~(\ref{first})
means that the operator has a 
nonvanishing projection onto the scalar hadronic states as well.
In the language of Green's functions
this means that the correlator 
\be
T_{\mu\nu,\al\beta}(q)=i\int 
dx e^{iqx}\langle Tj_{\mu\nu}(x) j_{\al\beta}(0)\rangle
\label{corr}
\ee
gets contributions not only from the tensor mesons with 
$J^{PC}=2^{++}$ but also from the states with quantum numbers
$J^{PC}=0^{++}$, or the scalar gluonium.
Therefore the correlator~(\ref{corr}) is not described by
the single scalar function when radiative corrections are included.
The most general tensor decomposition of the correlator~(\ref{corr}) 
has the form
\be
T_{\mu\nu,\al\beta}(q) = \eta_{\mu\nu,\al\beta}(q)\ T(q^2)
+f_{\mu\nu,\al\beta}(q)\ T_S(q^2)
\label{tendec}
\ee
where the tensor object $\eta_{\mu\nu,\al\beta}(q)$
is defined through the elementary transverse tensors
\[
\eta_{\mu\nu} = q_\mu q_\nu - q^2 g_{\mu\nu} 
\]
by the expression
\be
\eta_{\mu\nu,\al\beta}(q)=\eta_{\mu\al}\eta_{\nu\beta}  
+\eta_{\mu\beta}\eta_{\nu\al}-\frac{2}{3}\eta_{\mu\nu}\eta_{\al\beta} .
\label{t4}
\ee
The quantity 
$\eta_{\mu\nu,\al\beta}(q)$ 
is a density (polarization) matrix of a particle with spin 2 
which determines the structure of its propagator in momentum space.
It has the properties
\be
q^\mu\eta_{\mu\nu,\al\beta}(q)=0,
\quad q^\al\eta_{\mu\nu,\al\beta}(q)=0,
\quad \eta_{\mu\mu,\al\beta}(q)=0,
\quad \eta_{\mu\nu,\al\al}(q)=0 .
\label{proper}
\ee
Note that the last two relations 
in (\ref{proper})
are valid only in four-dimensional space-time.
The proper tensor in $D$-dimensional space-time, necessary for
computation within dimensional regularization, reads
\be
\eta^D_{\mu\nu,\al\beta}(q) = \eta_{\mu\al}\eta_{\nu\beta}  
+\eta_{\mu\beta}\eta_{\nu\al}-\frac{2}{D-1}\eta_{\mu\nu}\eta_{\al\beta} .
\label{dt}
\ee
It has a vanishing trace and is orthogonal to the quantity 
$f_{\mu\nu,\al\beta}(q)$.
However, both are equivalent to the order of perturbation theory
we work in the present paper,
and the difference between eq.~(\ref{t4})
and eq.~(\ref{dt}) is inessential.
The quantity 
$\eta_{\mu\nu,\al\beta}(q)$ is symmetric in both pairs of indices
$(\mu\nu)$ and $(\al\beta)$.
The second tensor object entering eq.~(\ref{tendec})
\[
f_{\mu\nu,\al\beta}(q)=\eta_{\mu\nu}\eta_{\al\beta}  
\]
is the tensor structure related to the contribution 
of scalar particles.
In a general $D$-dimensional space-time 
the tensors 
$f_{\mu\nu,\al\beta}(q)$ and $\eta_{\mu\nu,\al\beta}(q)$ are not orthogonal 
to each other. 
The perturbation theory expansion of the function $T_S(q^2)$ starts with 
the terms
of order $\als^2$ (the nonvanishing imaginary part)
and is negligible in the considered 
(next-to-leading) order in $\als$
to which we limit ourselves in the present paper.
Nonvanishing term of this tensorial structure 
emerges due to the trace anomaly.
Thus, up to the terms of order $O(\als^2)$ the correlator (\ref{corr}) 
is determined by the single scalar function $T(q^2)$
related to the contribution of tensor gluonia only. 
The anomalous dimension of the current $j_{\mu\nu}$
vanishes that makes the function $T(q^2)$
invariant with regard to the renormalization group transformations.

The results of direct computations 
of the function $T(q^2)$ are as follows.
The leading order contribution to 
the function $T(q^2)$ is well known \cite{SVZglu1}.
Within dimensional
regularization with $D=4-2\eps$ being the space-time dimensionality
it has the form 
\be
{N_c^2-1 \over (4\pi)^2}
\frac{1}{10\eps}\left(\frac{\mu^2}{Q^2}\right)^\eps
G(\eps), \quad
Q^2=-q^2 \ .
\ee
The quantity $G(\eps)$ is related to the particular definition
of the integration measure in $D$-dimensional momentum space-time
and has the series expansion $G(\eps)=1+O(\eps)$ at small $\eps$
\cite{Gscheme}, $\mu$ is the t'Hooft mass of dimensional regularization. 
The factor $N_c^2-1$ counts the number of gluons (or partons at this
level of computation within perturbation theory) 
of the color group $SU(N_c)$
propagating in the
single loop to which the correlator reduces in this order of
perturbation theory.

For the amplitude $T(Q^2)$ with account for the two-loop perturbation
theory corrections one finds
\be
T(Q^2)
={N_c^2-1 \over 10(4\pi)^2} \frac{1}{\eps}
\left(\frac{\mu^2}{Q^2}\right)^\eps G(\eps)
\left(1+\frac{\als}{4\pi}N_c\left(-\frac{10}{9}\right)
\left(\frac{\mu^2}{Q^2}\right)^\eps G(\eps)\right) \ .
\label{mainres}
\ee
This expression represents the main result of the present note.
The coefficient of the leading order correction to the correlator was
first computed in ref.~\cite{diss} and later confirmed by independent 
computation \cite{chet}.
The basic technique of the computation is described in detail in
ref.~\cite{scaglu} where the scalar and pseudoscalar cases were considered.
Diagrams are the same in the tensor case.
Explicit expressions for the vertices and results for 
individual
diagrams
can be found in ref.~\cite{diss}.
The only complication in the present case is the tensor structure of 
diagrams. There are several different tensor projectors
which can convert the necessary expressions into the scalar form.
Some related integrals can be found in 
ref.~\cite{spinn} where the correlator of 
quark currents with spin $n$ was
considered. 
Poles of order $\eps^2$ which are possible at the two-loop level 
(in the $\als$ order) and actually present in
the expressions for particular diagrams
cancel in expression~(\ref{mainres})
as it should be because of renormalization group invariance of the
current $j_{\mu\nu}$.
For the corresponding $D$-function (a derivative of the amplitude 
$T(Q^2)$) which is multiplicatively renormalized
one has 
\be
D_T(Q^2)=
-Q^2\frac{d}{d Q^2}T(Q^2)=
{N_c^2-1 \over 160\pi^2} 
\left(1-\frac{\als}{4\pi}N_c \frac{20}{9} \right).
\ee
The coefficient of $\als$ in the last expression 
is independent of the renormalization scheme used for the calculation of
the amplitude $T(Q^2)$;
it is also seen in the fact that one has no need to fix a precise
definition of the integration measure: the quantity 
$G(\eps)$ enters the final answer only as a
factor $G(0)=1$. 
Evaluating the numerical value for this coefficient is the real
content of two-loop calculations of the correlator~(\ref{corr})
and the amplitude $T(q^2)$ in eq.~(\ref{tendec}) in particular.

Numerically, for the standard gauge group with $N_c=3$ one has 
\be
D_T(Q^2)=
{1 \over 20\pi^2} \left(1-\frac{5}{3}\frac{\als(Q^2)}{\pi} \right)\ .
\label{tenglu}
\ee
Thus, the next-to-leading order correction to the correlator of tensor 
gluonic currents is not large and the perturbation theory expansion 
is quite well convergent 
numerically even at $\als \approx 0.3$ that corresponds to the scale of
ordinary hadrons. 
The remarkable feature of this correction is its sign. In most cases
$\als$ corrections are positive while for the tensor correlator 
of currents (\ref{first}) it is
negative.
Inclusion fermions has a two-fold effect: loop corrections through
the gluon propagator and the mixing with fermionic operators at the
tree level. The former effect is trivial 
(explicit results can be found in ref.~\cite{diss}) while the latter 
was discussed in the literature.  

The smallness of the first order correction
to the observables which are
renormalization group invariant at the parton level
is a rather general feature of hadron phenomenology. 
The most famous example is the total cross section of $e^+e^-$ annihilation
into hadrons where the perturbation theory 
corrections in $\msbar$-scheme are not large, 
with the leading order correction explicitly given by 
\[
\sigma_{tot}(e^+e^- \rightarrow hadrons)\sim 1+\frac{\als}{\pi}+\ldots
\]
However, in cases when corrections depend on the
definition of the coupling constant in the considered channels
(as in the case of the 
scalar or pseudoscalar gluonium) they can be rather large in 
$\msbar$-scheme.
For instance, for the pseudoscalar gluonium with the
interpolating operator 
\[
j_{PS}=\als G\tilde G 
= \als \frac{1}{2}\eps^{\mu\nu\al\beta} G^a_{\mu\nu} G^a_{\al\beta}
\]
one finds \cite{scaglu}
\be
D_ {PS}(Q^2)=
{2\als^2(Q^2) \over \pi^2} 
\left(1+\frac{97}{4}\frac{\als(Q^2)}{\pi} \right)\ .
\label{psglu}
\ee
The correction of order $\als$ in eq.~(\ref{psglu})
is much larger than that in eq.~(\ref{tenglu}) and makes perturbation
theory inapplicable at momenta of the order of the standard values 
of hadronic resonance masses. 

It is worth noticing that the $D$-functions are defined in Euclidean
domain while the physical spectrum requires the correlators to be
evaluated on the physical cut. For two-point correlators the analytic
properties are well established and the analytic continuation can be
done in all order in $\als$ (e.g. \cite{an}) which can, however, change
the effective 
numerical magnitude of the total correction in concrete applications. 

Whether the magnitude of corrections reflects
the physical situation -- contribution
of instantons and early breakdown of
perturbation theory -- is still an open question
which urgently requires further investigation. 

\subsection*{Acknowledgments}
The author thanks K.G.~Chetyrkin for discussion,
encouragement and correspondence.
The work is supported in part by the Volkswagen 
Foundation under contract No.~I/73611 and 
by the Russian Fund for Basic Research under contracts
Nos.~97-02-17065 and 99-01-00091.

\end{document}